# Optical diffraction from $Ge_2Sb_2Te_5$ fishnet metasurfaces

D.V. Bochek[a], D.A. Yavsin[b], A.B. Pevtsov[b], K.B. Samusev[a,b], M.F. Limonov[a,b],*

[a] Department of Physics and Engineering, ITMO University, 197101 St. Petersburg, Russia
[b] Ioffe Institute, St. Petersburg 194021, Russia.

**ABSTRACT**

We report the creation of $Ge_2Sb_2Te_5$ metasurfaces on sapphire substrates by the ablation method and the study of their structural properties by scanning electron microscopy (SEM), atomic force microscopy (AFM), and optical diffraction. The main emphasis is on the optical technique, which boils down to obtaining bright Laue diffraction patterns on a screen, observing them with the naked eye, and analyzing the fine structure of diffraction reflections. It has been demonstrated that in one simple optical experiment it is possible to assess the quality of fabricated metasurfaces, determine the structure symmetry, and, moreover, determine the number of structural elements and lattice constants of the micron-sized metasurface. The accuracy of the optical technique is confirmed by comparison with the results of studies by SEM and AFM methods.

## 1. Introduction

Chalcogenide phase-change alloys composed of germanium (Ge), antimony (Sb), and tellurium (Te), including the most commercialized compounds $Ge_2Sb_2Te_5$ and $Ge_3Sb_2Te_6$ are employed as the active metasurfaces and metadevices [1-6]. Phase change $Ge_2Sb_2Te_5$ (hereinafter GST) materials can be easily switched between disordered amorphous state and ordered crystalline state using a thermal quenching cycle. This phase switching changes the refractive index of GST from a value of $n_a = 3.9 + 4.2i$ in the amorphous phase to $n_c = 4.3 + 2.0i$ in the crystalline phase at a wavelength of 0.73 μm [7] and for $Ge_3Sb_2Te_6$ from $n_a \approx 3.5 + 0.001i$ to $n_c \approx 6.5 + 0.06i$ at 3.1μm [8]. Owing to these flexibilities, the GST - alloys can be used as the active materials in information storage devices and for tuning many properties, including absorption and emission [9,10] chirality [11], nanoantenna resonance characteristics [12-14], as a platform for wavefront switch that can operate over 500 nm bandwidth at near-infrared spectral bands [15] among others.

Optical metasurfaces, a recently emerging field of micro- and nano-optics, derive their properties from single-layer planar artificial structures. In this case, the control of optical waves based on metasurfaces includes only the interaction between light and the structure of subwavelength thickness, which leads to a significant decrease in optical loss. This advantage, as well as the ease of fabrication of a two-dimensional structure compared to bulk materials, makes metasurfaces a new platform for studying the light-matter interaction, using their ability to manipulate the fundamental properties of optical waves, including phase, amplitude, polarization, and angular momentum. Great research interest has been focused on phase change material metasurfaces, and a significant amount has focused on Ge-Sb-Te-based two-dimensional structures.

The optical properties of both three-dimensional metamaterials and metasurfaces based on GST, including transmission, reflection, absorption and Raman spectra, have been actively studied [12-24]. Transmission and absorption studies have focused on the mid-infrared region, where strong spectral changes were observed as a result of the phase transition between the amorphous and crystalline phases, which can be used for practical applications. A broad spectral shift of 500 nm in the mid-infrared region was experimentally demonstrated in the transmission mode for the tunable Au+$Ge_2Sb_2Te_5$ metasurface, and the results were in good agreement with finite-difference time-domain simulations of the same structure [14]. This transmissive metasurface can provide a simple collinear filtering detector geometry and is therefore promising for integrated sensor chips and hyperspectral imaging arrays in the mid-infrared region. In a recent paper [16], the authors claim to be demonstrating best-in-class phase-change tunable metasurface filter based on a GST -embedded plasmonic nanohole array with the highest transmission efficiency of ~ 70% and the narrowest bandwidth of ~ 74 nm within the 3–5 μm waveband, and near perfect reflection off-resonance. A phase change compound GST with high transmittance and polarization conversion rate has also been used to create a high performance phase plate [17]. In the amorphous phase of GST, the metasurface operates as a quarter-wave plate in the wavelength range of 10.0–11.9 μm, and after the transition to the crystalline state, the metasurface acts as a half-wave plate in the wavelength range of 10.3–10.9 μm. A nanostructured GST film on a fused silica substrate was optimized in the 1.55 μm wavelength range to switch from a high transmission regime of 80% to a regime with a high absorption rate of 76% in transmission regardless of polarization, when thermally transformed from an amorphous to a crystalline state [18]. Note also that the study of phase-change chalcogenide materials has also extended to nonlinear optics. The second harmonic generation response from the oriented GST grains, induced by polarized laser pulses, was experimentally demonstrated [19]. It was found that the orientation of the GST grains is perpendicular to the direction of polarization of the pump laser. In GST -coated opaline photonic crystals, light control mediated by interplay of Wood anomalies and 3D Bragg diffraction resonances was demonstrated [20, 21]. Raman spectroscopy measurements were performed to reveal the detailed phase state of GST [23]. We conclude our short review by mentioning that in phase-change materials based on GST, it is possible to produce laser-induced periodic ripples which are formed by interference of the incident laser beam with the surface electromagnetic wave excited by this beam [22, 24].

Analyzing the literature on optical studies, we did not find any works reporting on the investigation of phase-change chalcogenide structures by diffraction methods. However, optical diffraction is a powerful tool for studying the structural and optical properties of various photonic structures, both three-dimensional photonic crystals [25-27] and two-dimensional films [28,29]. The fine structure of Laue optical diffraction from two-dimensional periodic photonic lattices of various symmetries — square $C_{4v}$, orthogonal $C_{2v}$, and hexagonal $C_{6v}$ — were studied experimentally and theoretically [28]. The structures consisted of submicron dielectric elements fabricated by direct laser writing. A surprisingly strong optical diffraction from a finite number of dielectric submicron elements was observed. Thus, Laue diffraction turns out to be an excellent tool for determining the symmetry, shape, and exact number of particles in a structure with a limited number of submicron scatterers.

In this study, we fabricated GST films of 200 μm thickness on a $Al_2O_3$ substrate, then, using the direct laser writing technique [30, 31], we created GST metasurfaces in the form of periodic arrays of square islets with different levels of ordering. Further, the samples were experimentally investigated by two methods - optical diffraction and SEM.

## 2. Metasurfaces design and fabrication

In this work we used the original technique of laser electrodispersion (LED) to fabricate nanostructured amorphous GST films (see details in [32]) onto silica or sapphire substrates. The technique is based on the process of cascade fission of molten submicrometer droplets splashed from the target surface by an intense laser pulse. The uniqueness of the LED technique is that the substrate is not involved in the process of nanoparticle formation. Therefore, neither the type of the substrate material nor the surface roughness affects the structural parameters of the deposited films. The technique enables us to control the density of surface coverage with nanoparticles.

Disks (diameter 12 mm, thickness 5 mm) made of polycrystalline GST, synthesized from special-purity elements Ge, Sb, and Te by melt-quenching in conical cells, were used as targets. The Nd:YAG laser we used had a wavelength of 1064 nm, a pulse repetition rate of 30 Hz, and a focused spot diameter of 1 mm. The laser power density of $10^9$ W/cm$^2$ was sufficiently for effective melting and evaporation of the target material and optical breakdown of the vapor to give a laser torch. As a result, under electrostatic forces in the spatial region of the laser torch, the fission of micrometer and submicron droplets up to nanometer size occurs. The resulting nanodroplets are rapidly cooled to become amorphous nanoparticles and are deposited onto a substrate.

X-ray fluorescence microanalysis confirmed that the composition of the films deposited onto the substrate corresponds to that of the GST compound. X-ray diffraction analysis demonstrated that the chalcogenide films are in the amorphous state. Transmission electron microscope (TEM) showed that the deposited films are composed of GST nanoparticles with sizes of 2 to 5 nm.

The thickness of the GST films depended on the deposition time (i.e., on the number of laser pulses) and could vary in our experiments from 10 to 200 nm. Independent measurements of the film thickness were conducted using an Ambios Technology XP-1 profilometer.

Direct laser writing technic was used to fabricate metasurfaces from a GST film on a sapphire substrate. We used a train of femtosecond pulses centered at around 780 nm wavelength and at a repetition frequency of 80 MHz. These pulses are derived from a 50-fs TiF-100F laser (Avesta-Project, Russia). Laser radiation was focused in the film through the glass substrate with a 40x microscope objective with numerical aperture NA = 0.75 (Carl Zeiss MicroImaging GmbH, Jena, Germany). The sample was mounted on a two-coordinate motorized linear air-bearing translator (Aerotech Inc., Lexington, KY, USA). The motion of the translators was controlled by a computer with software developed at Laser nanoFab GmbH, Germany.

The metasurfaces were created by direct fs-laser ablation of the GST film. The modes of removal of GST stripes from the substrate surface were investigated, which ensure complete removal of material and a clear shape of the metasurface elements that are obtained as a result of this process. The speed of the translators was varied in the range of 50- 150 µm/s, the laser power was varied in the range of 80- 130 mW. The time to create one well-ordered metasurface of about 100 x 100 µm$^2$ in size and the number of square elements of the order of 30 was approximately 2 minutes.

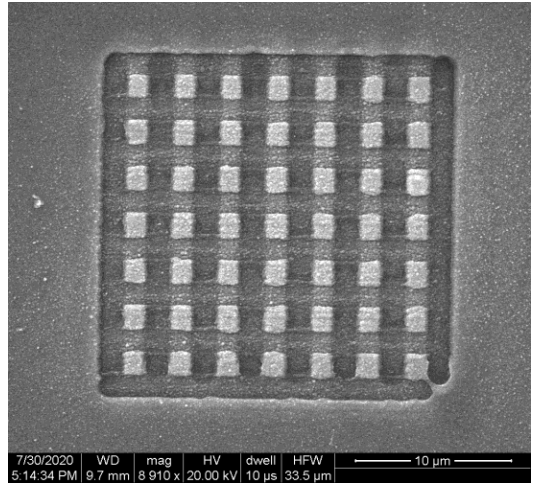

Fig. 1. SEM images of the 3 μm period fishnet metasurfaces on sapphire substrate.

## 3. Laue diffraction from 2D square structures

For the 2D structures with the square lattice symmetry (Fig.1), the position of each scatterer is determined by the 2D vector $\mathbf{r}_i = \mathbf{a}_1 n_1 + \mathbf{a}_2 n_2$, where $\mathbf{a}_1$ and $\mathbf{a}_2$ stand for the basis mutually perpendicular vectors ($\mathbf{a}_1 \cdot \mathbf{a}_2 = 0$) of the square lattice (lattice constants $a_1 = a_2$ in our case), $0 \le n_j \le N_{1,2}$, $N_{1,2}$ is the number of elements along the directions $\mathbf{a}_{1,2}$ respectively, and $n_j$ are integer.

To analyze 2D diffraction patterns, it is sufficient to use the Born approximation [33], when only single scattering is considered. If all elements have the same shape, the diffraction intensity is determined by the square of the structure factor:

$$|S(\mathbf{q})|^2 = \frac{\sin^2(N_1 \mathbf{q}\mathbf{a}_1/2)}{\sin^2(\mathbf{q}\mathbf{a}_1/2)} \frac{\sin^2(N_2 \mathbf{q}\mathbf{a}_2/2)}{\sin^2(\mathbf{q}\mathbf{a}_2/2)}. \quad (1)$$

Here $\mathbf{q} \equiv \mathbf{k}_i - \mathbf{k}_s$ is the scattering vector, whereas $\mathbf{k}_i$ and $\mathbf{k}_s$ are the wave vectors of the incident and diffracted waves. To analyze the diffraction patterns, we consider scattering by a one-dimensional linear chain of scatterers; for this, in the equation (1) we set $N_2 = 1$. The positions of the strong diffraction maxima are determined in the limit $\sin(\mathbf{q}\mathbf{a}_1/2) \to 0$, which leads to the expression $\mathbf{q}\mathbf{a}_1 = (\mathbf{k}_i - \mathbf{k}_s)\mathbf{a}_1 = 2\pi n$, which determines the diffraction condition. For normal incidence of light $\mathbf{k}_i \mathbf{a}_1 = 0$, the equation $\mathbf{k}_s \mathbf{a}_1 = 2\pi n$ takes the form $\frac{2\pi a_1}{\lambda} \cos\theta_s = 2\pi n$ describing a plane perpendicular to the chain axis and a family of pairs of cones with symmetry axes coinciding with $\mathbf{a}_1$, and apex angle of scattering is given by the expression:

$$\theta_s = \arccos(n\lambda/a_1), \quad (2)$$

where $n$ is an integer denoting the diffraction order, $\theta_s$ is the scattering angle between vectors $\mathbf{a}_1$ and $\mathbf{k}_s$. The experiment shows that in the case of a sample with a square lattice on a screen located perpendicular to the laser beam, the Laue diffraction pattern is a simple superposition of patterns from two systems of mutually perpendicular chains. Figure 2 shows the simulation of Laue diffraction and diffraction patterns on the screen in different projections for a sample with mutually perpendicular scattering chains. The diffraction pattern with the $C_{4v}$ square symmetry consisting from two orthogonal strips that correspond to the zero-order scattering and four arcs that correspond to the first-order scattering.

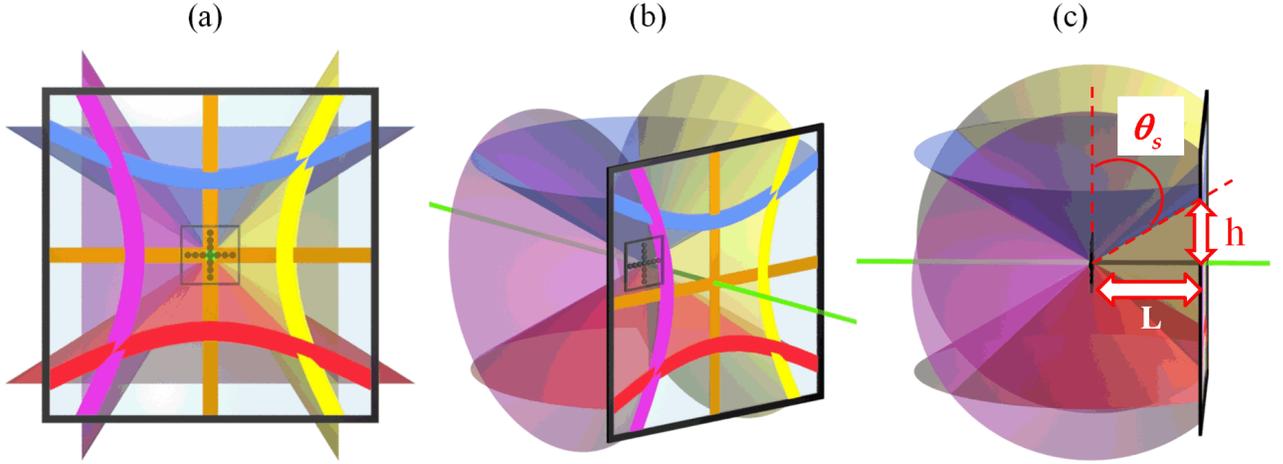

Fig. 2. Schematic of the zero-order (n = 0) and first-order (n = 1) Laue diffraction from the square structure (the horizontally and vertically oriented chains of scatterers) in the case $a_1 = a_2$. Diffraction patterns on a flat screen are shown by thick lines. Three images from different angles: rear view of the screen (a), at an angle to the screen (b) and perpendicular to the screen (c). On (c) the distances from the sample to the screen L, the distance on the screen from the laser beam to the first diffraction arch h and the angle at the apex of the scattering cone $\theta_s$ are indicated. Different components of the scattered light are shown in different colors for clarity.

## 4. Results and discussion

Before describing experiments on optical diffraction, we will focus on an important terminological point. As is known, 3D metamaterials are considered to be effective media with averaged parameters of electric and magnetic susceptibility and lattice constants much smaller than the length of electromagnetic waves $a \ll \lambda$. As follows from expression (2), in this case only zero-order Bragg (or Laue) diffraction can be observed. However, in the case of metasurfaces, the situation is different, since the expected effects are observed when the phase of the electromagnetic wave changes at a length comparable to the lattice constant (see, for example, [34]). Therefore, for our optical experiments, we choose the excitation laser wavelength ($\lambda$ = 0.532 μm) less than the lattice parameter of the metasurface $a > \lambda$ in order to study the Laue diffraction of various orders, as is observed in the case of photonic crystals [28].

For optical diffraction experiments, a glass substrate with a thickness of 180 μm (size about 1 cm$^2$) with a set of metasurfaces was mounted on a goniometric slide, which is usually used in X-ray diffraction measurements and allows the sample to be aligned with high accuracy about three spatial axes. We used a second harmonic Nd:YAG laser with a wavelength $\lambda$ = 0.532 μm. Optical diffraction patterns were observed in the backscattering geometry on a flat screen located between the sample and the laser perpendicular to the laser beam; photographs were taken with an Olympus C-2040 Zoom.

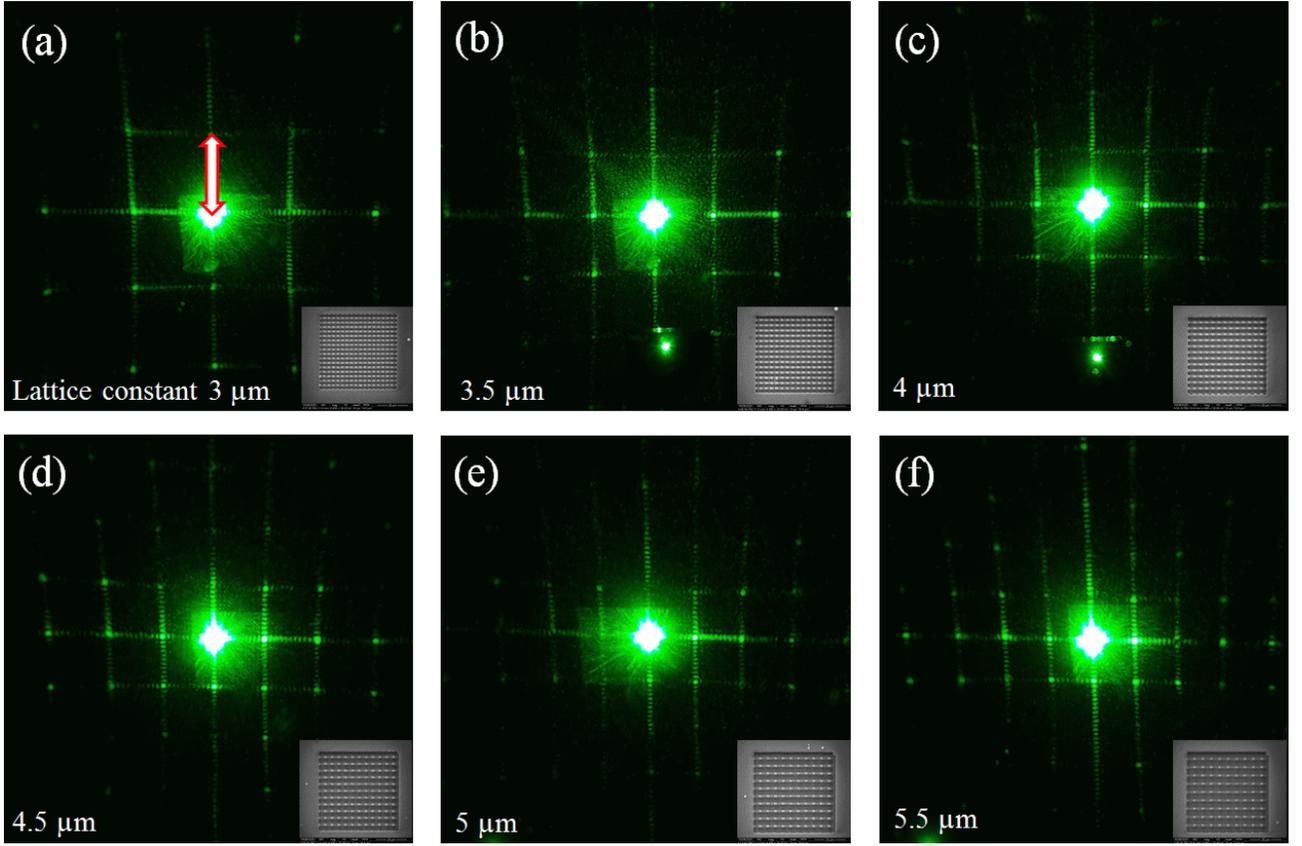

Fig. 3 Experimentally measured diffraction patterns obtained from GST metasurfaces with different lattice constants $a_1 = a_2$. The insets show SEM images of the corresponding structures. $\lambda = 0.532$ μm.

From equation (2) it follows that at $a < \lambda$ all scattering is reduced to the zero-order cone ($n = 0$) that degenerates into a plane normal to $\mathbf{a}$, since the angle between and $\mathbf{k}_s$ and $\mathbf{a}$ is equal to $\theta_s = \pi/2$. The first pair of diffraction cones appears at $a > \lambda$ and further pair of $n$-th order cones appears when $a > n\lambda$, but is forbidden when the argument is outside the arccosine region $|n\lambda/a| > 1$. Thus, as the parameter $a$ increases, the number $n$ of observed diffraction cones increases, while the angles at their apex $\theta_s$ decrease. Figure 3 presents diffraction patterns experimentally observed from a set of metasurfaces with variable lattice parameter 3μm ≤ $a_1 = a_2$ ≤ 5.5μm. It clearly demonstrates the behavior of diffraction patterns depending on the lattice parameter $a$. With increasing $a$, the number of paired arcs observed on the screen increases and, accordingly, the difference $\theta_{s,n} - \theta_{s,n+1}$ decreases, which is reflected on the screen as the corresponding diffraction arcs coming closer together.

It is important to note that only two measurements of macroscopic dimensions make it possible to determine the lattice constant of the studied metasurface with high accuracy. These dimensions are the distance from the sample to the screen (Fig.2c, in our experiment it was L=20.7 cm) and the distance on the screen from the center of the laser beam to the first diffraction arc, marked with a red arrow in Fig. 2c and Fig.3a, in that case was h=3.65 cm. Simple calculations allow us to determine the angle at the top of the diffraction cone as $\theta_s = 80°01'$, after which, using equation (2), taking into account that $n = 1$ and $\lambda = 0.532$ μm, we obtain the value of the lattice constant $a = 3.07$ µm. Thus, the error between the values obtained from the optical experiment and from the SEM data is about 1.5%, which can be considered as a good result, taking into account the simplicity of the diffraction experimental setup.

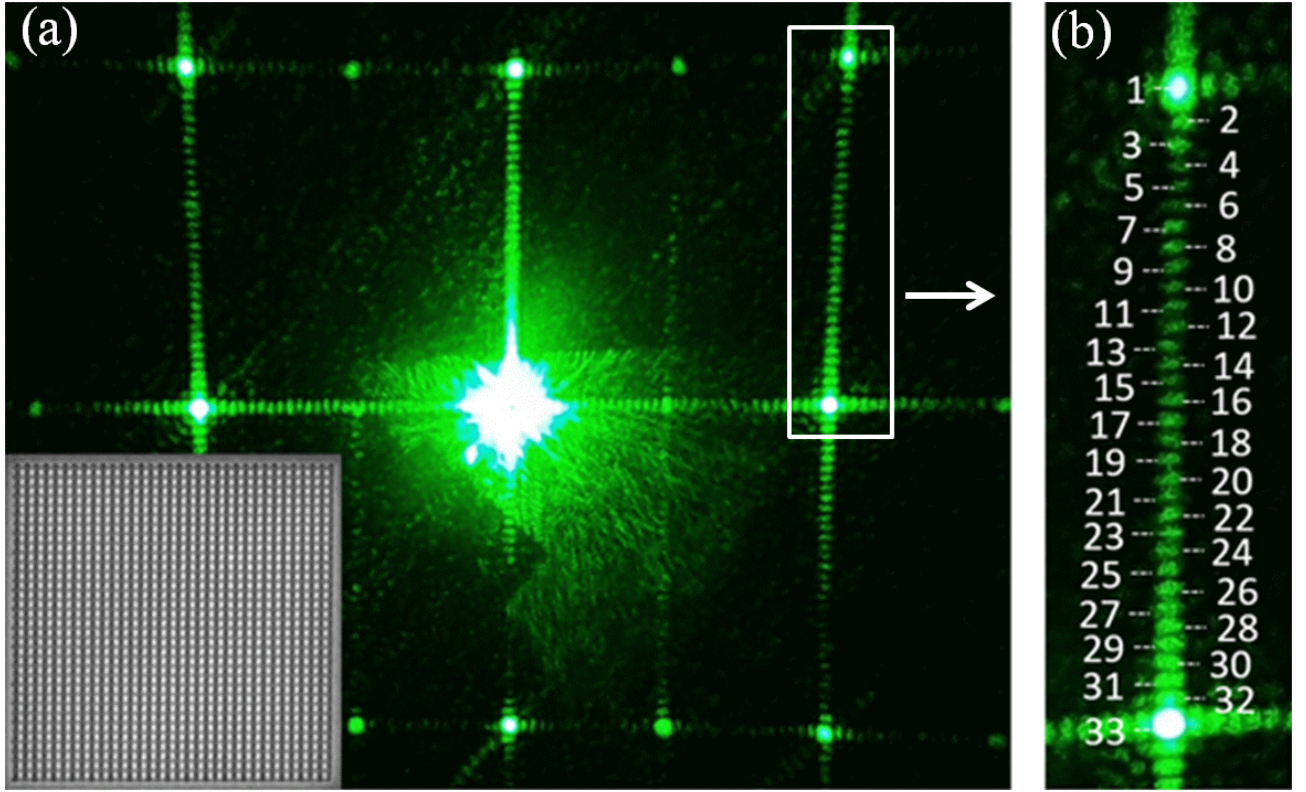

Fig. 4 (a) Experimentally measured diffraction pattern obtained from 100x100 µm GST metasurface with period of 3 µm. Inset shows the SEM image of the structure. (b) The enlarged area of the diffraction arc, marked on (a) with a white rectangle. The picture shows a fine structure consisting of 33 diffraction maxima. $\lambda = 0.532$ µm.

Next, we analyze a fine structure of the diffraction planes and cones, which is observed in the experimental diffraction patterns shown in Fig. 4. The origin of this structure follows from the equation (1) for the structure factor $S(\mathbf{q})$. First, two main intense maxima in the diffraction patterns (both on planes, $n = 0$, and on cones, $n = \pm 1, \pm 2,...$) are determined by the vanishing of the denominator of the structure factor $\sin(\mathbf{qa}_1/2) = 0$. Further, the numerator $\sin(N_1\mathbf{qa}_1/2)$ in the equation (1) for $S(\mathbf{q})$ has $N_1$-1 zeros between any two adjacent main maximums and, accordingly, $N_1$-2 additional maximums. Thus, together with the two main intense scattering maxima that arise due to the vanishing of the denominator, we obtain the total number of maxima $N_1$, i.e. for an ordered square metasurface, we can determine the number of elements $N_1$ and $N_2$ independently along each axis $\mathbf{a}_1$ and $\mathbf{a}_2$. Therefore, with the naked eye, it is possible to determine the number of micron and sub-micron elements of the metasurface using the diffraction pattern on a screen. In particular, from Fig. 4 (b) it can be seen that the number of diffraction maxima between the two main ones is 31, including 2 main maxima, we obtain 33, which exactly corresponds to the number of metasurface elements along the corresponding (in this case, the horizontal) axis.

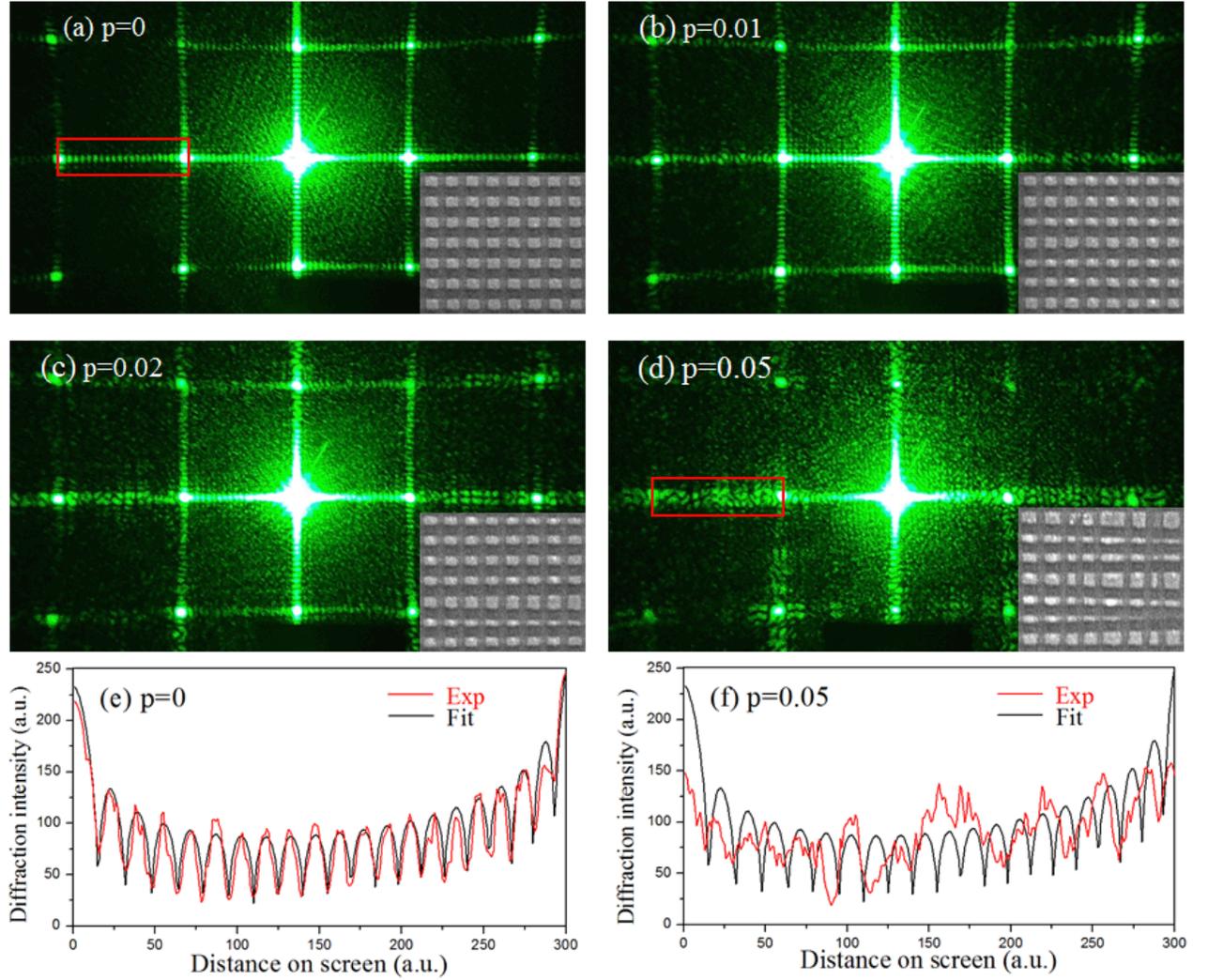

Fig.5. (a-d) Evolution of the diffraction patterns during the transition from ordered (a) to disordered (b-d) metasurfaces. Insets: fragments of SEM images of the corresponding metasurfaces with disorder parameters $p = 0$ (a), 0.01 (b), 0.02 (c), 0.05 (d). (e,f) Experimental intensity of the diffraction stripes marked in (a) and (e) with red rectangles for $p = 0$ (e) and $p = 0.05$ (f) (red curves), and the intensity calculated using equation (1) (black curves). $\lambda = 0.532$ μm.

Finally, we investigated the influence of structural disorder of metasurfaces on the Laue diffraction patterns. In these experiments, we continue the study of glassy metasurfaces, which we started earlier [29], while demonstrating a new method for processing experimental results. Using the laser ablation method, two types of metasurfaces were created: with an ideally ordered arrangement of square GST islands (Fig.5a) and a random shape and size of GST islands (Fig. 5b-d). We fabricated the disordered metasurfaces as follows. Each pass of the laser beam, removing GST stripes from the substrate surface, was oriented around its central point (along the $x$ or $y$ axis in the $xy$ plane) at a random angle $\alpha_i$ with respect to the ordered state ($\alpha_i = 0$ for the rods in the ordered structure). To create a set of disordered metasurfaces, we employed uniform distribution function:

$-\alpha_{max} \leq \alpha_i \leq \alpha_{max}$ where $\alpha_{max} = p\frac{\pi}{4}, 0 \leq p \leq 1$. It is clear that the GST islands on the disordered metasurfaces no longer had a rectangular shape, as on the ordered ones. The lattice constant of the ordered metasurface was 3 μm.

The results of experimental studies of Laue diffraction from ordered and disordered fishnet metasurfaces depending on the disorder parameter *p* are presented in Fig. 5, together with fragments of SEM imagines of the corresponding samples. To analyze the relationship between optical diffraction patterns and the magnitude of the structural disorder of the metasurfaces, we chose a region along the zero-order stripe between the first and second main maxima. These regions (red rectangles in Fig. 5) are not significantly affected by Rayleigh scattering observed on the screen around the laser beam. In the case of an ordered metasurface (p = 0), all the fine structure maxima determined by the number of scatterers of the metasurface are clearly distinguished on the strip. Even with a slight increase in the structural disorder, which is difficult to distinguish in the SEM image (inset in Fig. 5b), the fine structure begins to noticeably blur, completely disappearing at p = 0.05 (Fig. 5d).

The experimental data were processed according to equation (1) taking into account the background scattering and the geometry of the screen, which is flat and not spherical. In the case of an ordered metasurface, a close to ideal agreement between the calculation and experiment is observed (Fig 5e). The insignificant inaccuracy is due to a number of factors, such as fluctuations in laser power, inhomogeneity of the original GST film. For a disordered film (p = 0.05), the maxima of the fine structure are hardly traced or completely absent. Statistical analysis shows that the MSD parameters (Mean Square Deviation) for p = 0 and p = 0.05 differ by about three times.

## 5. Conclusions

In this article, we have demonstrated the powerful capabilities of a simple and generally available optical diffraction method for characterizing micron-scale metasurfaces with structural elements several microns in size. Using GST metasurfaces fabricated by ablation on sapphire substrates as experimental samples, we have shown that diffraction patterns can be observed on a screen in reflection geometry with the naked eye and that information about the metasurface quality and the exact number of structural elements that form the metasurface can be obtained. Simple measurements of the geometrical dimensions of the setup and diffraction patterns on the screen with a ruler with a millimeter scale make it possible to determine with high accuracy the micron and sub-micron lattice constants of the metasurface without using the scanning electron microscopy technique. The proposed method for determining the parameters of metasurfaces can be used for a variety of samples on transparent (in the transmission geometry) and opaque (in the reflection geometry) substrates.

**Financial support**

This work was supported by the Russian Science Foundation (Project 20-12-00272).

**Declaration of competing interest**

The authors declare that they have no known competing financial interests or personal relationships that could have appeared to influence the work reported in this paper.

## Acknowledgments

The authors are grateful to A.D. Sinelnik for help in the experiments and to P.A. Belov and M.V. Rybin for discussing the results.
## References

1. Q. Wang, E.T.F. Rogers, B. Gholipour, C.-M. Wang, G. Yuan, J. Teng, N.I. Zheludev, Optically reconfigurable metasurfaces and photonic devices based on phase change materials, Nature Photon. 10 (2016) 60.
2. P. Li, X. Yang, T.W.W. Maß, J. Hanss, M. Lewin, A.-K.U. Michel, M. Wuttig, T. Taubner, Reversible optical switching of highly confined phonon–polaritons with an ultrathin phase-change material, Nature Mater. 15 (2016) 870.
3. L. Kang, R.P. Jenkins, D.H. Werner, K. J. Vahala, Recent progress in active optical metasurfaces, Adv. Optical Mater. 7 (2019) 1801813.
4. T. Cui, B. Bai, H.-B. Sun, Tunable metasurfaces based on active materials, Adv. Funct. Mater. 29 (2019) 1806692.
5. S. Chen, Z. Li, W. Liu, H. Cheng, J. Tian, From Single-Dimensional to Multidimensional Manipulation of Optical Waves with Metasurfaces, Adv. Mater. 31 (2019) 1802458.
6. A.V. Kolobov, J. Tominaga, P. Fons, Phase-change memory materials, in: S. Kasap, P. Capper (eds), Springer Handbook of Electronic and Photonic Materials, Springer Handbooks, Springer, Cham, 2017.
7. K. Shportko, S. Kremers, M. Woda, D. Lencer, J. Robertson, M. Wuttig, Resonant bonding in crystalline phase-change materials, Nature Mater. 7 (2008) 653.
8. X. Yin, T. Steinle, L. Huang, T. Taubner, M. Wuttig, T. Zentgraf, H. Giessen, Beam switching and bifocal zoom lensing using active plasmonic metasurfaces, Light: Sci. Appl. 6 (2017) e17016.
9. A. Tittl, A.-K. U. Michel, M. Schäferling, X. Yin, B. Gholipour, L. Cui, M. Wuttig, T. Taubner, F. Neubrech, H. Giessen, A switchable mid-infrared plasmonic perfect absorber with multispectral thermal imaging capability, Adv. Mater. 27 (2015) 4597.
10. T. Cao, X. Zhang, W. Dong, L. Lu, X. Zhou, X. Zhuang, J. Deng, X. Cheng, G. Li, R.E. Simpson, Tuneable thermal emission using chalcogenide metasurface, Adv. Opt. Mater. 6 (2018) 1800169.
11. X. Yin, M. Schäferling, A.-K.U. Michel, A. Tittl, M. Wuttig, T. Taubner, H. Giessen, Active chiral plasmonics, Nano Lett., 15 (2015) 4255.